\documentclass[prl,aps,twocolumn,a4paper,floatfix,showpacs]{revtex4}

\usepackage{graphicx,psfrag}
\usepackage{mathrsfs}
\usepackage{amsmath,amsfonts,amssymb}
\usepackage{multirow}
\usepackage{comment,hyperref}

\newcommand{\be}{\begin{equation}}
\newcommand{\ee}{\end{equation}}
\newcommand{\bea}{\begin{eqnarray}}
\newcommand{\eea}{\end{eqnarray}}
\newcommand{\bel}{\begin{align}}
\newcommand{\eel}{\end{align}}

\newcommand{\TEOBGSF}{{TEOB$_{\rm Resum}$}}
\newcommand{\TEOBNNLO}{{TEOB$_{\rm NNLO}$}}

\usepackage{color}

\definecolor{cyan}{rgb}{0,0.9,0.9}
\definecolor{orange}{rgb}{0.9,0.5,0}
\definecolor{magenta}{rgb}{1,0,1}
\definecolor{purple}{rgb}{0.8,0.4,0.8}
\definecolor{gray}{rgb}{0.8242,0.8242,0.8242}

\begin{document}

\title{Modeling the dynamics of tidally-interacting binary neutron
  stars up to merger}  

\author{Sebastiano \surname{Bernuzzi}$^{1,2}$}
\author{Alessandro \surname{Nagar}$^3$}
\author{Tim \surname{Dietrich}$^4$}
\author{Thibault \surname{Damour}$^3$}
\affiliation{$^1$TAPIR, California Institute of Technology, Pasadena,
  California, USA} 
\affiliation{$^2$DiFeST, University of Parma, I-43124
  Parma, Italy} 
\affiliation{$^3$Institut des Hautes Etudes Scientifiques, 91440
  Bures-sur-Yvette, France}  
\affiliation{$^4$Theoretical Physics Institute, University of Jena,
  07743 Jena, Germany}

\date{\today}

\begin{abstract}
  The data analysis of the gravitational wave signals emitted by
  coalescing neutron star binaries requires the availability of an
  accurate analytical representation of the dynamics and waveforms of
  these systems. 
  We propose an effective-one-body (EOB) model that describes the
  general relativistic dynamics of neutron star binaries from the
  early inspiral {\it up to merger}. 
  Our EOB model incorporates an enhanced attractive tidal potential
  motivated by recent analytical advances in the post-Newtonian
  and gravitational self-force description of relativistic tidal
  interactions. No fitting parameters 
  are introduced for the description of tidal interaction in the late, 
  strong-field dynamics.
  We compare the model energetics and the gravitational wave phasing  
  with new high-resolution multi-orbit numerical relativity simulations 
  of equal-mass configurations with different equations of state.  
  We find agreement within the uncertainty of the numerical data 
  for all configurations. 
  Our model is the first semi-analytical model which captures the tidal
  amplification effects close to merger. It thereby provides the most 
  accurate analytical representation of binary neutron star dynamics 
  and waveforms currently available. 
\end{abstract}


\maketitle

\paragraph{Introduction.---}

One of the key aims of the upcoming detections of gravitational wave (GW) 
signals from coalescing  binary neutron stars (BNS) is to inform us 
on the equation of state (EOS) of matter at supranuclear
densities~\cite{Read:2009yp,Damour:2012yf,Read:2013zra,DelPozzo:2013ala,Lackey:2014fwa}
via the measurement of the 
tidal polarizability coefficients (or Love
numbers)~\cite{Flanagan:2007ix,Hinderer:2007mb,Damour:2009vw,Binnington:2009bb,Hinderer:2009ca}
that enter both the interaction potential 
and the waveform. A necessary requirement for this program is the
availability of faithful waveform models that capture  the
strong-gravity and tidally-dominated regime of the late-inspiral 
of BNS {\it up to merger}. Such models are presently missing;
the aim of this work is to close this gap so as to help 
developing GW astronomy.    

The theoretical modeling of BNS waveforms is challenging, and requires 
synergy between analytical and numerical approaches to 
the general relativistic two body problem.
Traditional post-Newtonian (PN) analytical methods reach their
limits during the late BNS inspiral, and are a major
limitation for GW data
analysis~\cite{Favata:2013rwa,Yagi:2013baa,Lackey:2014fwa}.  
In recent years numerical relativity (NR) simulations have become fairly
robust~\cite{Baiotti:2010xh,Bernuzzi:2011aq,Bernuzzi:2012ci,Hotokezaka:2013mm,Radice:2013hxh,Bernuzzi:2013rza}, 
though the achievable precision is under debate and exploring the
physical parameter space at the necessary accuracy (waveform length
and phase errors) is certainly out of
reach~\cite{Bernuzzi:2011aq,Hotokezaka:2013mm,Radice:2013hxh}. 
The difficulties related to PN and NR modeling carry over in the
construction of hybrid PN-NR templates~\cite{Read:2013zra}. 
Presently, the effective-one-body (EOB) 
formalism~\cite{Buonanno:1998gg,Buonanno:2000ef,Damour:2000we,Damour:2001tu} 
offers the most accurate analytical description of the relativistic two body 
problem.
By combining information coming both from
analytical results and numerical simulations,
the EOB framework succeeds in describing the 
energetics and the GW signals of coalescing and merging
black hole binaries
(BBH)~\cite{Damour:2012ky,Pan:2013rra,Pan:2013tva,Taracchini:2013rva,Damour:2014afa,Damour:2014yha,Damour:2014sva}.

\begin{figure}[t]
  \begin{center}
    \includegraphics[width=0.45\textwidth]{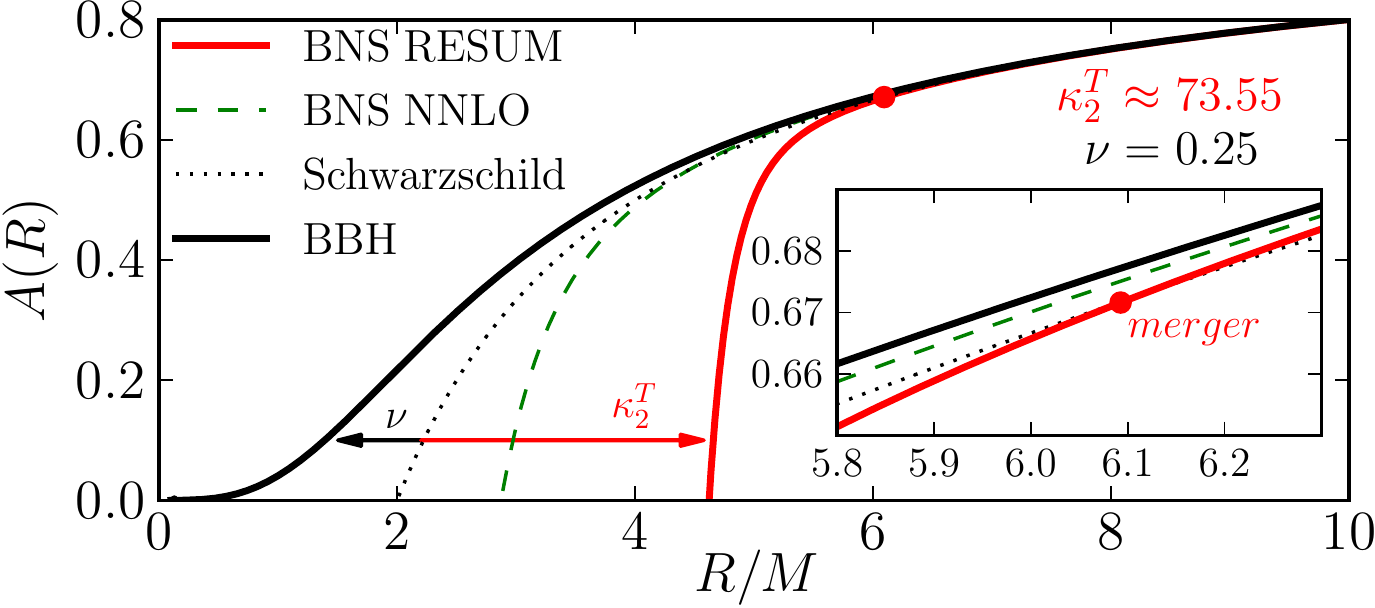}
    \caption{%
      The main radial gravitational potential $A(R)$ in various
      EOB models.
      Finite-mass ratio effects ($\nu$) make the 
      gravitational interaction less attractive than the Schwarzschild
      relativistic potential $A_{\rm Schw}=1-2M/R$, while
      tides ($\kappa_2^T$, see Table~\ref{tab:configs}) 
      make it more attractive (especially at short separations).}
     \label{fig:Apotential}
   \end{center}
 \end{figure}

 The EOB model is a relativistic generalization of the well-known
 Newtonian property that the relative motion of a two-body system
 is equivalent to the motion of a particle of mass $\mu=M_A M_B/(M_A+M_B)$
 in the two-body potential $V(R)$. The Newtonian radial
 dynamics is governed by the effective potential $V_{\rm eff}(R;\,P_\varphi)=P_\varphi^2/(2\mu R^2)+V(R)$,
 where the first term, which contains the angular momentum $P_\varphi$, is a centrifugal potential.
 In the EOB formalism there is an analogous 
 effective relativistic radial potential (setting $G=c=1$),
 $W_{\rm eff}(R;\,P_\varphi) =\sqrt{A(R)\left(\mu^2 + (P_\varphi/R)^2\right)}$,
 where $A(R)$ is the main radial potential. 
 In the Newtonian approximation, $A(R)\approx1 + 2 V(R)/\mu$,
 so that $W_{\rm eff}(R)\approx \mu + V_{\rm eff}(R;\,P_\varphi)$. 
 In the test-mass limit, $A(R)$ is  simply equal to the Schwarzschild 
 potential $A_{\rm Schw}=1-2M/R$ (where $M\equiv M_A+M_B$). 
 Beyond the test-mass limit, 
 $A(R)$ is a deformation of  $A_{\rm Schw}$ by two different
 physical effects: (i) finite-mass ratio effects, 
 parametrized by $\nu\equiv \mu/M$; and (ii) tidal effects 
 (in BNS systems only), 
 parametrized by relativistic tidal polarizability parameters 
 $\kappa_A^{(\ell)}$~\cite{Hinderer:2007mb,Damour:2009vw,Binnington:2009bb,Hinderer:2009ca},
 the most important of which is the quadrupolar combination $\kappa_2^T=\kappa_A^{(2)}+\kappa_B^{(2)}$.
 Following~\cite{Damour:2009wj,Damour:2012yf}, tidal interactions are
 incorporated in the EOB formalism by a radial potential of the form
 $A(R;\,\nu;\,\kappa_A^{(\ell)})=A^0(R;\,\nu)+A^T(R;\kappa_A^{(\ell)})$ 
 where $A^0(R)$ is the EOB BBH
 radial potential, and $A^T(R)$ is an additional tidal interaction 
 piece whose structure is discussed below.
 Figure~\ref{fig:Apotential} contrasts 
 the deformations of $A(R;\,\nu;\,\kappa_A^{(\ell)})$
 away from $A(R;0;0)=A_{\rm Schw}$ induced 
 either by (i)~finite-mass-ratio effects, which make $A^{0}(R)$
 less attractive, or by (ii)~tidal effects, which make $A^{\rm BNS}(R)$
 more attractive in the strong-field regime where they dominate over
 the repulsive finite-mass-ratio effects.  
 Figure~\ref{fig:Apotential} also compares a {\it resummed} tidal EOB model 
 (incorporating recent advances in the relativistic theory of tidal 
 interactions~\cite{Bini:2013rfa,Dolan:2014pja,Bini:2014zxa})
 with another tidal EOB model that incorporates a tidal potential 
 treating tidal interactions in a {\it nonresummed way}, up to the 
 next-to-next leading order (NNLO, fractional 2PN, see below)~\cite{Vines:2010ca,Bini:2012gu}.
 The resummed tidal EOB model is significantly more attractive than 
 the NNLO one at small separations.
 We will consider the evolution of the EOB dynamics at separations
 of the order of the contact between the two NSs, i.e.,
 at the point hereafter called ``merger''. The marker in the figure 
 indicates the radial location corresponding to that merger for the
 resummed EOB model ($R_{\rm mrg}=6.093M$). 

 The main result of this paper is to show that the resummed 
 EOB model is significantly closer (especially near merger) 
 to the results of new, high-resolution, multi-orbit NR simulations,
 than both the NNLO EOB model and the conventional 
 T4 PN model. This breaks new ground with respect 
 to previous EOBNR comparisons~\cite{Baiotti:2010xh,Bernuzzi:2012ci,Hotokezaka:2013mm}
 which could never display good analytical-numerical agreement 
 up to merger~\cite{Damour:2009vw,Baiotti:2010xh,Bernuzzi:2012ci,Hotokezaka:2013mm},
 and offers the first hope of analytically modeling BNS up to merger.

 \begin{table*}[t]
   \caption{BNS configurations and phasing results. 
     From left to right: name, EOS, $\kappa_2^T$, \TEOBNNLO~light-ring location,
     star compactnesses ${\cal C}_{A,B}$ and gravitational  
     masses in isolation, initial Arnowitt-Deser-Misner (ADM) mass
     and angular momentum, $(M_{\rm  ADM}^0,{\cal J}_{\rm ADM}^0)$. 
     The phase differences  
     $\Delta\phi^{\rm X}\equiv \phi^{\rm X}-\phi^{\rm NR}$, where
     $X={\rm TT4},\,{\rm TEOB_{\rm NNLO}},\,{\rm TEOB_{\rm
         Resum}})$ 
     labels various analytical models, are reported at the
     moment of NR merger  
     ($0.11\lesssim M\omega_{22}^{\rm mrg}\lesssim 0.19$) as well
     as the corresponding NR uncertainty $\delta\phi^{\rm NR}_{\rm NRmrg}$.
     The resummed \TEOBGSF~model displays the best
     agreement with NR data. The phase differences, in radians, 
     are obtained by aligning all waveforms on the 
      frequency interval $I_\omega\approx (0.04,0.06)$.} 
  \centering  
\begin{ruledtabular}
  \begin{tabular}{cccccccccc|cc}        
    Name & EOS & $\kappa_2^T$ & $r_{\rm LR}$ & ${\cal C}_{A,B}$ &
    $M_{A,B}[M_\odot]$ & $M_{\rm ADM}^0[M_\odot]$ &
    ${\cal J}_{\rm ADM}^0[M_\odot^2]$ & 
    $\Delta\phi^{\rm TT4}_{\rm NRmrg}$ & 
    $\Delta\phi^{\rm TEOB_{\rm NNLO}}_{\rm NRmrg}$ & 
    $\Delta\phi^{\rm TEOB_{\rm Resum}}_{\rm NRmrg}$ &
    $\delta\phi^{\rm NR}_{\rm NRmrg}$ \\  
   \hline
   % ----------------------------------------------------
   2B135 & 2B   & 23.9121 & 3.253 & 0.2049  & 1.34997 & 2.67762  
   & 7.66256
   %& 0.038  
   %&  0.127 & 1520 
   & $-1.25$ & $-0.19$ & $+0.57$\footnote{
This value is the dephasing at the moment of \TEOBGSF~merger, which
occurs $\approx 30M$ before NR merger after alignment.} & $\pm$4.20\\ 
   % ----------------------------------------------------
   SLy135 & SLy  & 73.5450 & 3.701 & 0.17381 & 1.35000 & 2.67760
   & 7.65780  
   %& 0.038
   %&  0.107 & 1280  
   & $-2.75$ & $-1.79$ & $-0.75$  & $\pm$0.40\\ 
   % ----------------------------------------------------
   $\Gamma_2164$&$\Gamma=2$ & 75.0671 & 3.728  & 0.15999 & 1.64388 & 3.25902
   &11.11313  
   %& 0.0414     
   %& 0.107 & 1052 
   & $-2.29$ & $-1.36$ & $-0.31$ & $\pm$0.90\\ 
   % ----------------------------------------------------
   $\Gamma_2151$&$\Gamma=2$ &  183.3911 & 4.160 & 0.13999 & 1.51484 & 3.00497
   &9.71561
   %& 0.0367     
   %& 0.092 & 981 
   & $-2.60$ & $-1.92$ & $-1.27$ & $\pm$1.20\\ 
   % ----------------------------------------------------
   H4135 &H4   &  210.5866 & 4.211 & 0.14710 & 1.35003 & 2.67768 
   &7.66315
   %& 0.038  
   %& 0.090 & 1077 
   & $-3.02$  & $-2.43$ & $-1.88$ &$\pm$1.04\\ 
   % ----------------------------------------------------
   MS1b135 & MS1b &  289.8034 & 4.381 & 0.14218 & 1.35001 & 2.67769
   & 7.66517
   %& 0.038  
   %& 0.085 & 1017 
   &$-3.25$ & $-2.84$ & $-2.45$ &$\pm$3.01\\ 
  \end{tabular}
\end{ruledtabular}
  \label{tab:configs}
\end{table*}

\begin{figure*}[t]
  \begin{center}
    \includegraphics[width=0.325\textwidth]{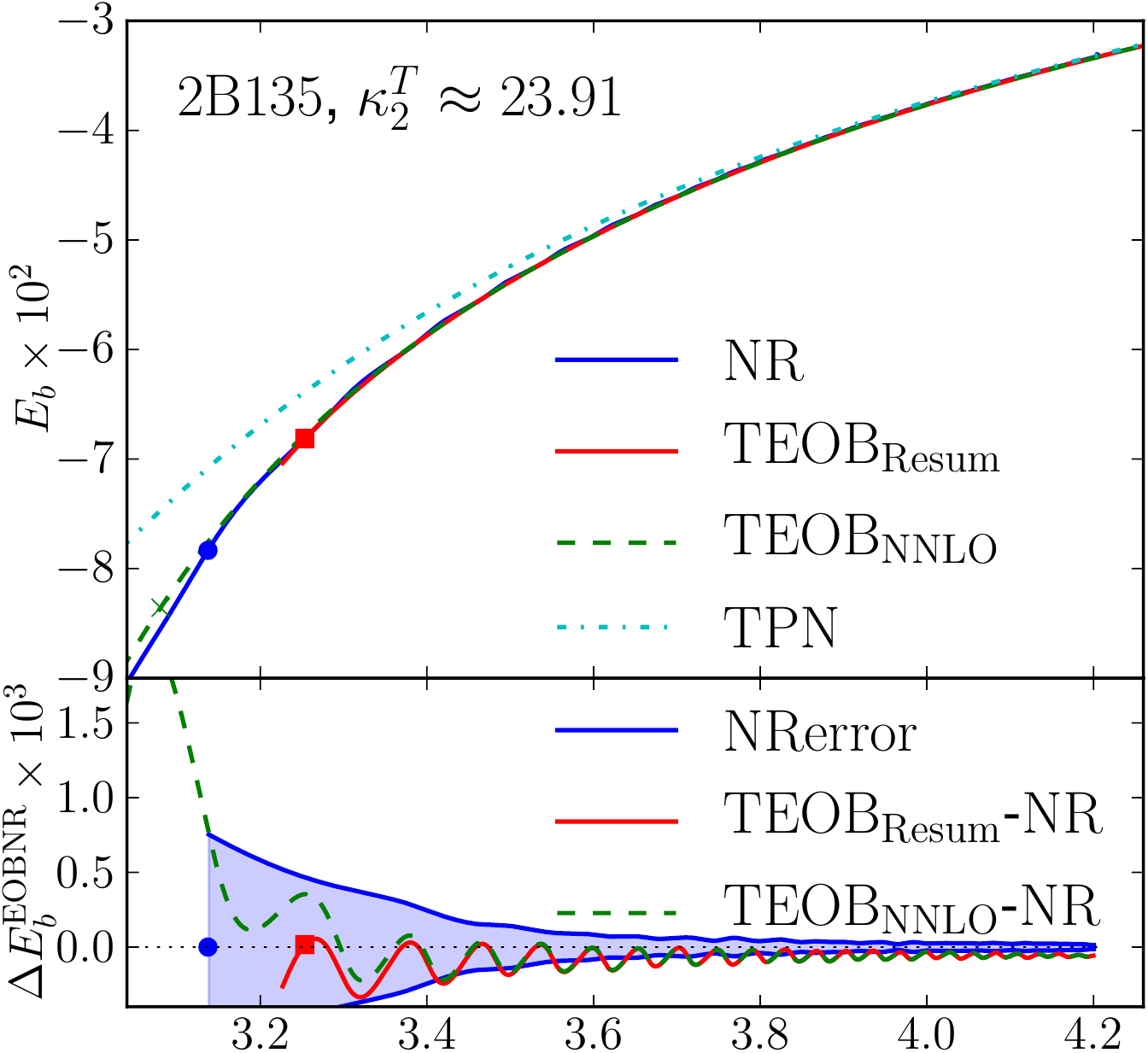}
    \includegraphics[width=0.325\textwidth]{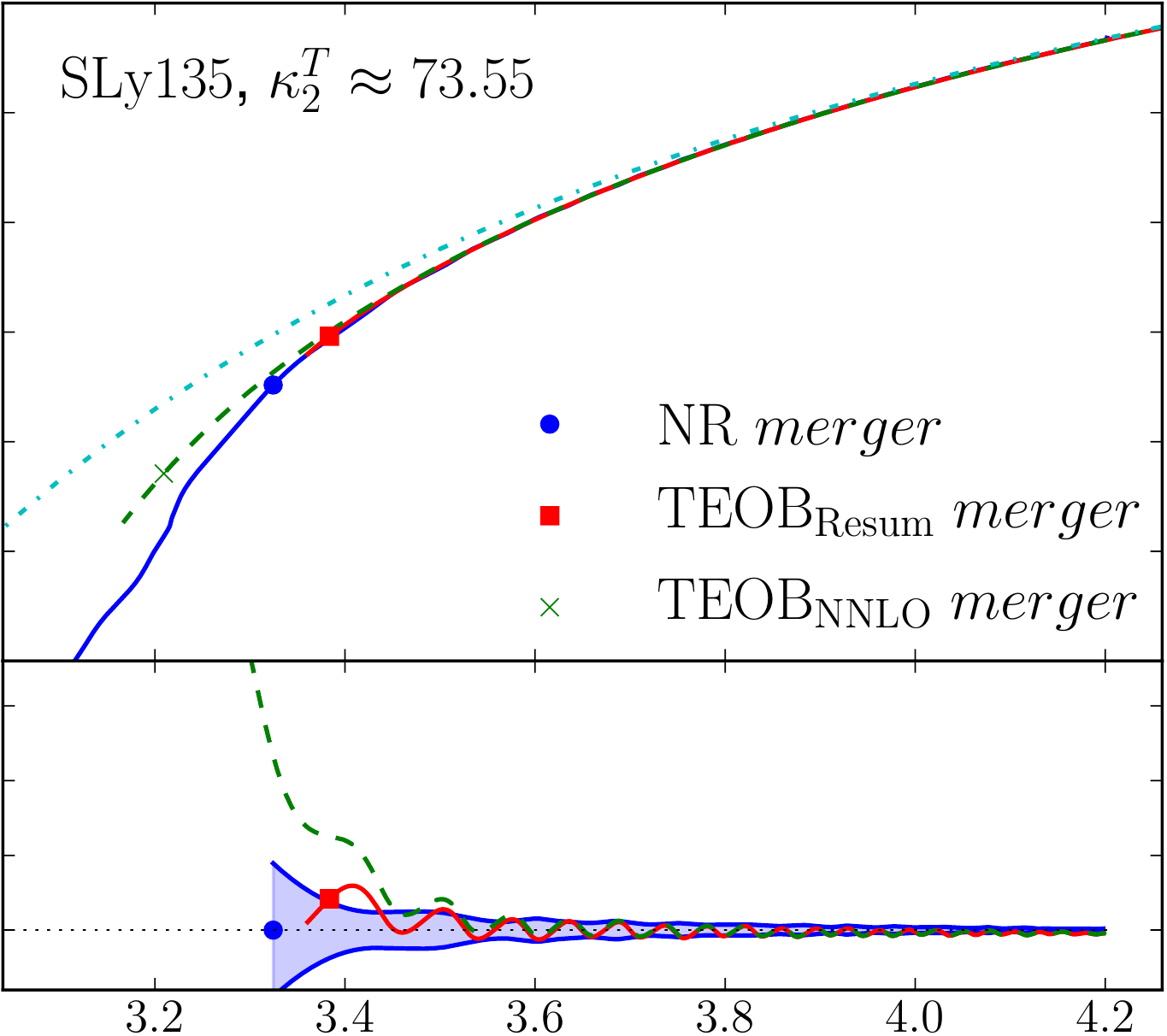}
    \includegraphics[width=0.325\textwidth]{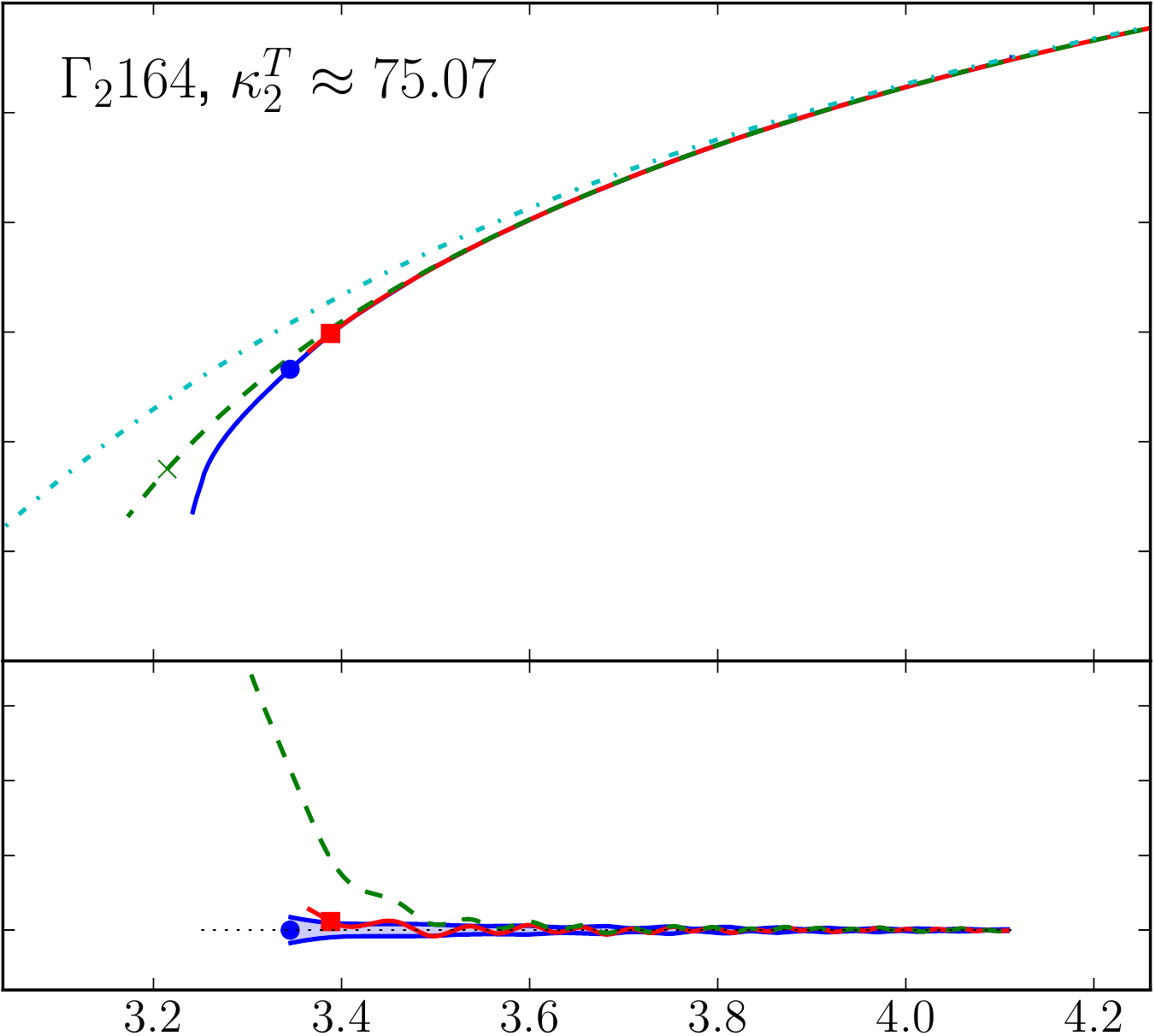}\\
    \includegraphics[width=0.325\textwidth]{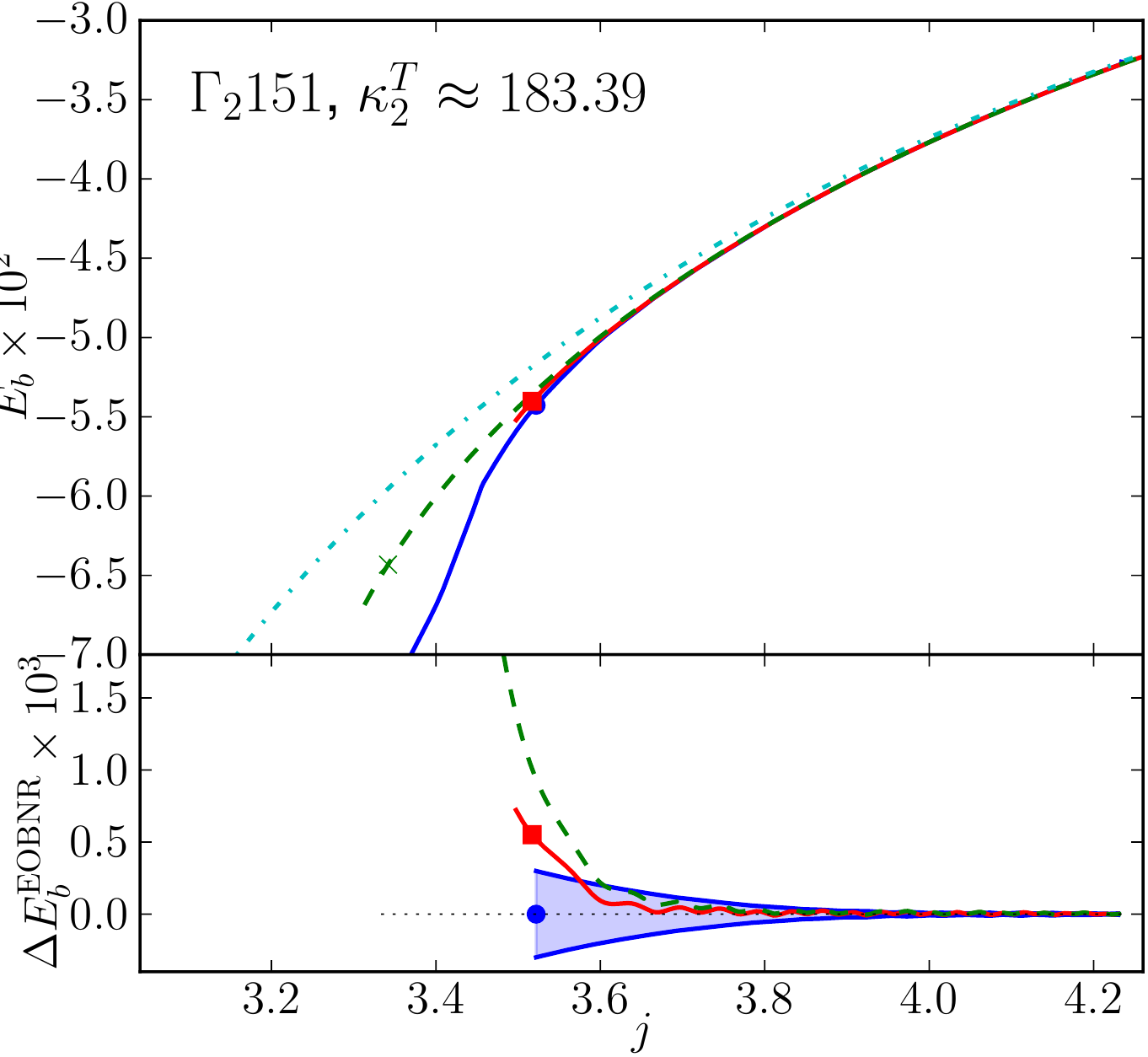}
    \includegraphics[width=0.325\textwidth]{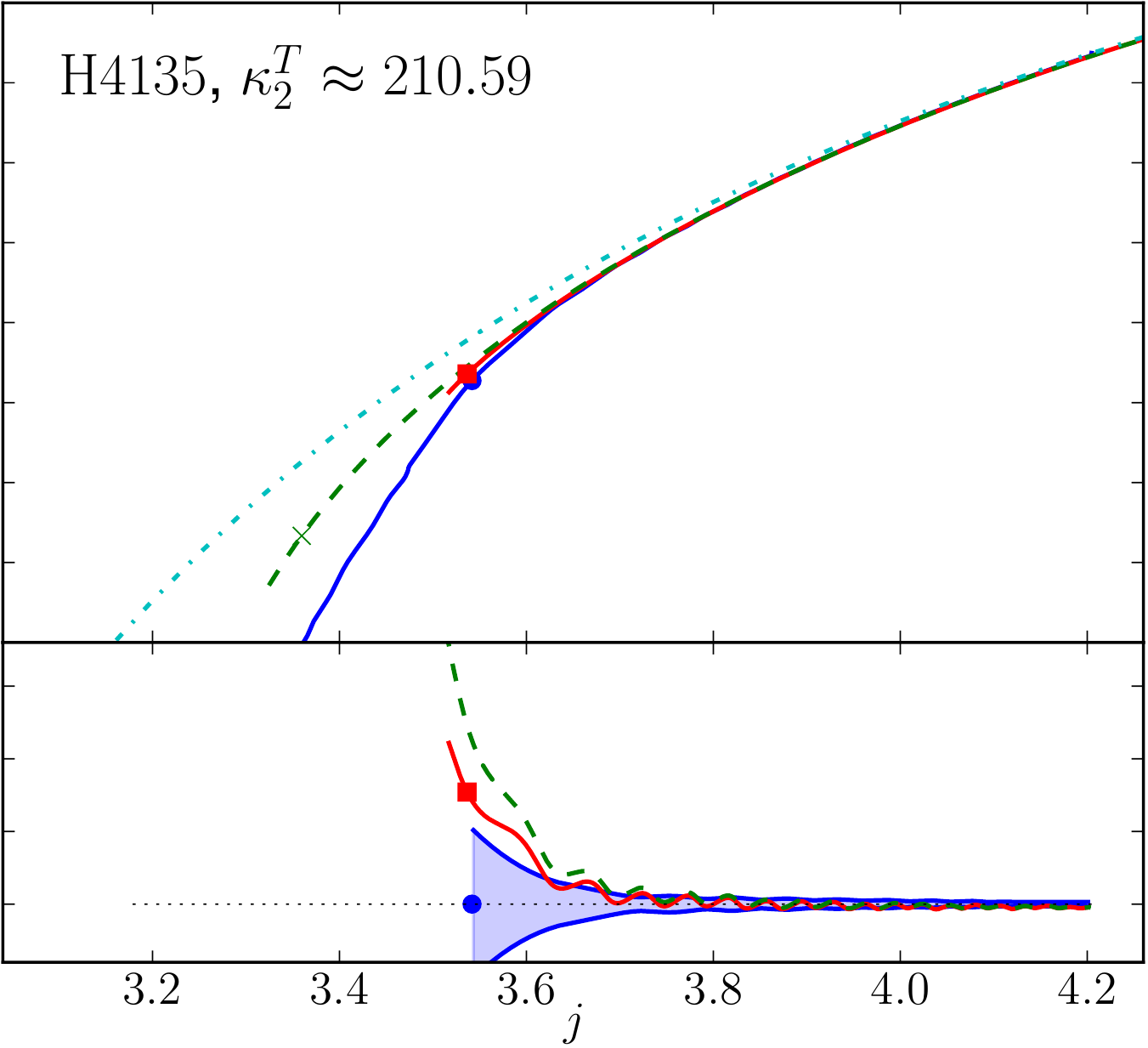}
    \includegraphics[width=0.325\textwidth]{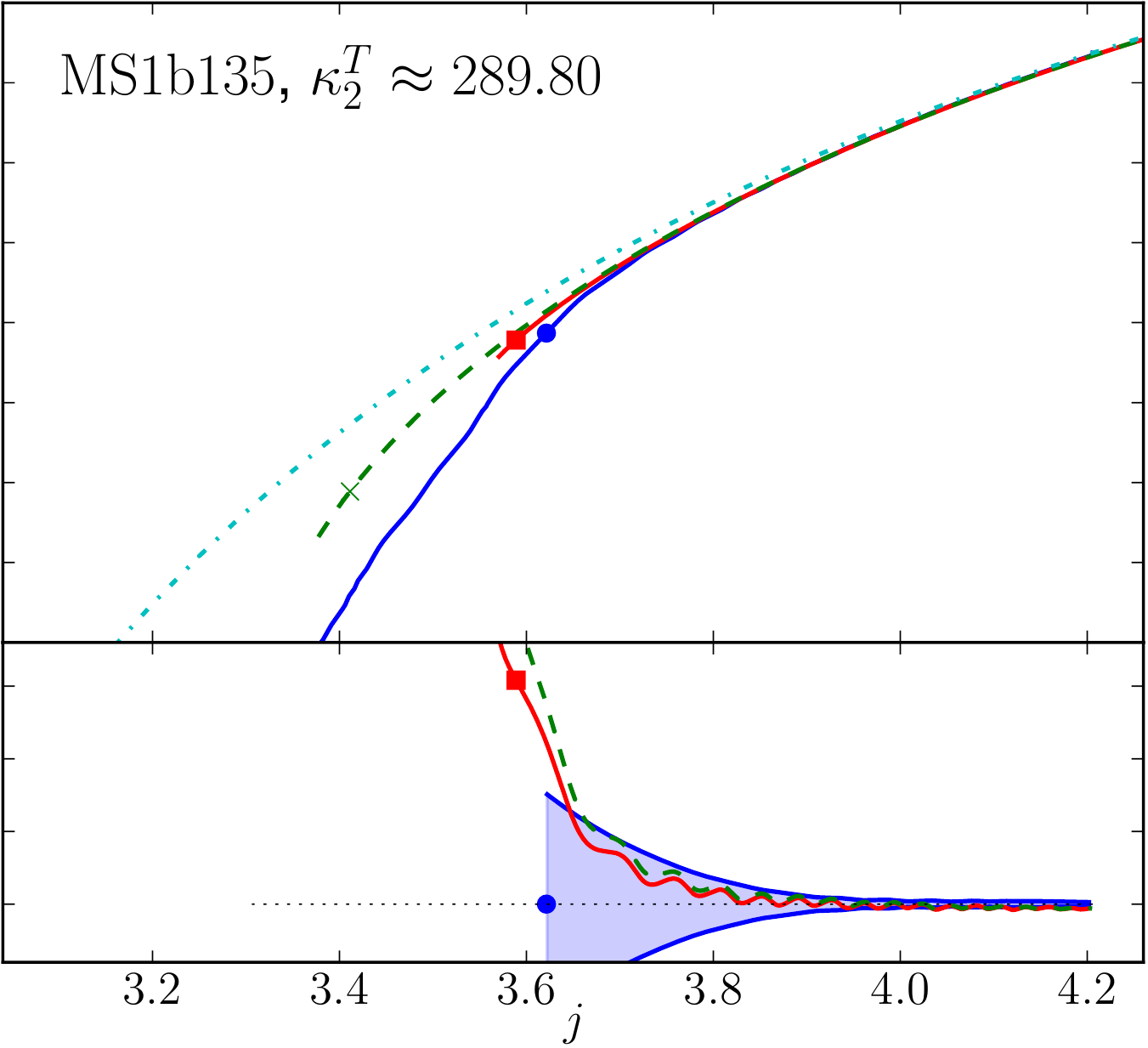}
    \caption{ 
      Energetics: comparison between NR data,
      \TEOBGSF, \TEOBNNLO~and TPN.  Each bottom panel shows the two 
      EOB-NR differences. The filled circles locate the merger points 
      (top) and the corresponding differences (bottom). 
      The shaded area indicates the NR uncertainty. The \TEOBGSF~model
      displays, globally, the smallest discrepancy with NR data (notably for merger quantities), 
      supporting the theoretical, light-ring driven, amplification of
      the relativistic tidal factor.}  
  \label{fig:dynamics}
  \end{center}
\end{figure*}

\paragraph{The tidal EOB models (TEOB).---}
The EOB Hamiltonian is $H_{\rm EOB} = M\sqrt{1+2\nu(\hat{H}_{\rm eff}-1)}$
where, in the nonspinning case, 
$ \hat{H}_{\rm eff}(u,p_{r_*},p_\varphi)\equiv H_{\rm eff}/\mu = 
\sqrt{A(u;\nu)\,(1  + p^2_\varphi u^2 + 2\nu(4-3\nu)u^2 p_{r*}^4) + p_{r*}^2} $, 
with $u\equiv 1/r\equiv GM/(R c^{2})$, $p_\varphi
\equiv P_{\varphi}/(M\mu)$, $p_{r_*}\equiv
\sqrt{A/B}p_r=P_{r_*}/\mu$, and where 
$A(u;\nu)\equiv A^0(u;\nu)+A^T(u;\nu)$ and $B(u;\,\nu)$ are EOB potentials.  
We define the BBH potential $A^{0}(u;\nu)$  as the $(1,5)$ 
Pad\'e approximant of the formal 5PN expression
$A^{0}_{\rm 5PN}(u;\nu)=1-2u +a_3 u^{3} + a_4 u^4 +
(a_{5}^{c}(\nu)+a_{5}^{\ln}\ln u) u^{5} +
(a_{6}^{c}(\nu)+a_{6}^{\ln}\ln u) u^{6}$. 
The coefficients up to 4PN, i.e.~$(a_3,a_4,a_5^c(\nu),a_5^{\ln})$, are
analytically known~\cite{Bini:2013zaa}.  
At 5PN, both $a_6^{\ln}$ and the linear-in-$\nu$ part
of $a_{6}^{c}(\nu)$~\footnote{ % 
The linear-in-$\nu$ corrections are actually completely known up to
8.5PN order~\cite{Bini:2014nfa}}  
are analytically known~\cite{Barausse:2011dq,Bini:2013rfa}. 
We do not use here the analytical knowledge of $a_{6}^{c}(\nu)$.
We used instead the ``effective'' value $a_6^c(\nu) =3097.3\nu^2 - 1330.6\nu + 81.38$
deduced from a recent comparison between the EOB model 
and a sample of NR data~\cite{Damour:2015prep,Mroue:2013xna}.
The tidal contribution to $A(r)$ (omitting the negligible 
gravitomagnetic part~\cite{Damour:2009vw}) is
\be
\label{AT}
A_T^{(+)}(u;\nu)\equiv - \sum_{\ell=2}^{4} \left[ \kappa^{(\ell)}_A
  u^{2\ell+2}\hat{A}^{(\ell^+)}_A + (A\leftrightarrow B)\right],
\ee 
where $\kappa^{(\ell)}_A = 2 k^{\ell}_A \left(X_A/{\cal
  C}_A\right)^{2\ell+1}M_B/M_A$, $X_{A,B}\equiv M_{A,B}/M$,  
$k^{(\ell)}_{A,B}$ are the dimensionless Love
numbers~\cite{Hinderer:2007mb,Damour:2009vw,Binnington:2009bb,Hinderer:2009ca}
and ${\cal C}_{A,B}\equiv (M/R_*)_{A,B}$ the stars compactnesses with 
$R_{*A,B}$ the areal radii. 
In the equal-mass case, 
the EOS information is essentially encoded in the total dimensionless
{\it quadrupolar} tidal coupling constant 
$\kappa^T_2 \equiv \kappa^{(2)}_A + \kappa^{(2)}_B$. 
The relativistic correction factors $\hat{A}^{(\ell^+)}_A$ formally include 
all the high PN corrections to the leading-order. The choice of 
$\hat{A}^{(\ell^+)}_A$ defines the two tidal EOB models of this paper.
The NNLO tidal EOB model, \TEOBNNLO, is defined by using the PN-expanded,
fractionally 2PN accurate, expression
$\hat{A}_A^{(\ell^+ ) \rm NNLO}= 1+ \alpha^{(\ell)}_1 u+ \alpha^{(\ell)}_2 u^2$
with $\alpha^{(2),(3)}_{1,2}\neq 0$ and $\alpha^{(4)}_{1,2}=0$~\cite{Bini:2012gu}. 
The resummed tidal EOB model, \TEOBGSF, is defined 
by using for the $\ell=2$ term in Eq.~\eqref{AT} the expression
\begin{align}
  \label{hatA2}
  \hat{A}^{(2^+)}_A(u) &= 1 + \dfrac{3u^2}{1-r_{\rm LR} u} 
  + \dfrac{X_A \tilde{A}_1^{(2^+) \rm 1SF}}{(1-r_{\rm LR} u)^{7/2}} %\nonumber\\ 
    + \dfrac{X_A^2\tilde{A}_2^{(2^+) \rm 2SF}}{\left(1-r_{\rm
      LR}u\right)^{p}}, 
\end{align}
where the functions $\tilde{A}_1^{(2^+)\rm 1SF}(u)$ and
$\tilde{A}_2^{(2^+)\rm 2SF}(u)$ are defined as
in~\cite{Bini:2014zxa}, and where we choose $p=4$ for the exponent.
The $\ell=3,4$ contributions of the resummed model are taken 
as in the NNLO model.
A key prescription here is to use as pole location in Eq.~\eqref{hatA2}
the light ring $r_{\rm LR}(\nu;\kappa_A^{(\ell)})\,$ (i.e., the 
location of the maximum of  $A^{\rm NNLO}(r;\,\nu;\,\kappa_A^{(\ell)})/r^2$) 
of the NNLO tidal EOB model~\cite{Bini:2012gu}. 
The radial part of radiation reaction, ${\cal F}_r=0$, is always
set to zero~\cite{Damour:2014sva,Damour:2015prep}; the tidal part 
of radiation reaction is completed with the next-to-leading-order tidal
contribution~\cite{Damour:2009wj,Vines:2010ca,Damour:2012yf}.

\paragraph{NR simulations.---}
Simulations are performed with the BAM
code~\cite{Brugmann:2008zz,Thierfelder:2011yi}, 
which solves the Z4c formulation of 
Einstein's equations~\cite{Bernuzzi:2009ex,Hilditch:2012fp} and
general relativistic hydrodynamics. 
The setup used here is similar to that
of~\cite{Bernuzzi:2012ci,Bernuzzi:2014kca}, 
numerical details will be discussed elsewhere. 
We consider equal-mass binaries in which the fluid is  
described either by a $\Gamma=2$ polytropic EOS enforcing isentropic
evolutions~\cite{Bernuzzi:2011aq,Bernuzzi:2012ci}, 
or by a piecewise polytropic representation of cold
EOS~\cite{Read:2008iy} adding a $\Gamma_{\rm th}=1.75$ 
thermal pressure component~\cite{Hotokezaka:2013mm}.
All configurations (Table~\ref{tab:configs}) are simulated 
at multiple resolutions. The simulations of 
(SLy,\,$\Gamma_2151$,\,H4) use three resolutions with 
$(64^3,\,96^3,\,128^3)$ grid points resolving 
the star diameter, while for 
(2B,\,$\Gamma_2164$,\,MS1b) only the $(64^3,96^3)$ resolutions are
available. 
Numerical uncertainties are conservatively estimated as the
difference between the highest and the second highest available
resolutions, in an attempt at including possible systematic
errors~\cite{Bernuzzi:2012ci}. 
Overall, these BNS data are among the longest and most accurate
available to date.

\begin{figure*}[t]
  \begin{center}
    \includegraphics[width=.32\textwidth]{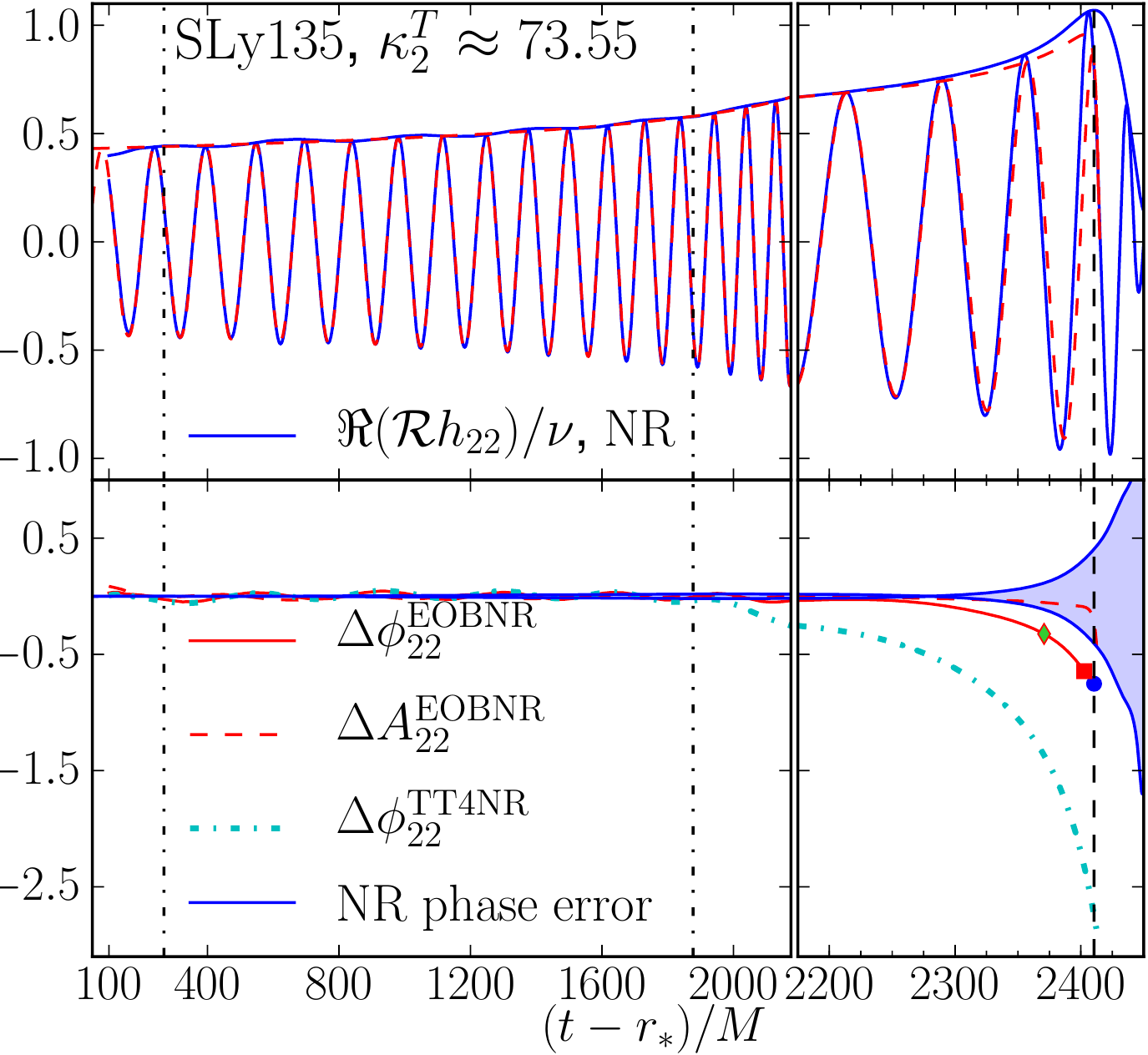}
    \includegraphics[width=.32\textwidth]{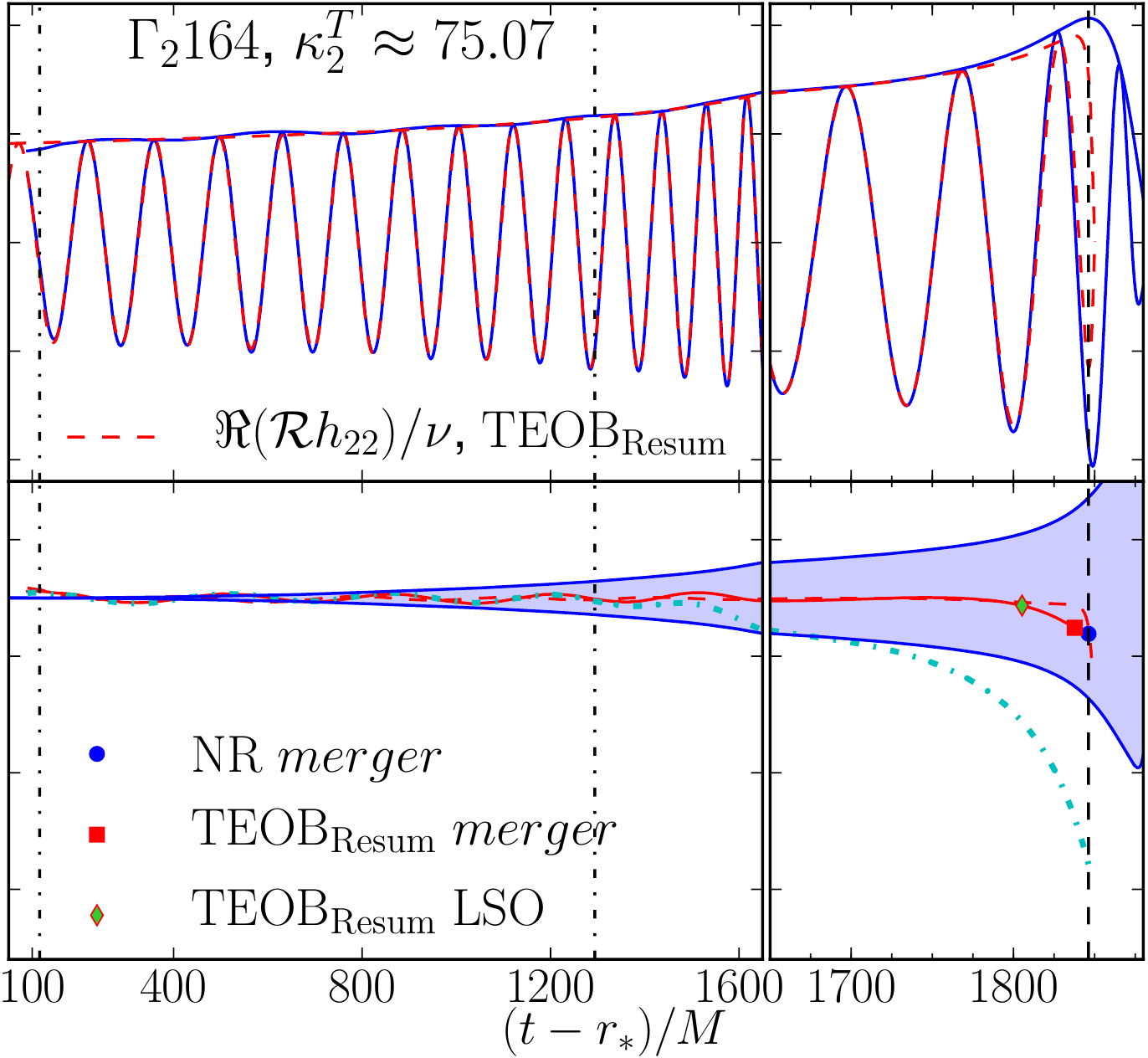}
    \includegraphics[width=.32\textwidth]{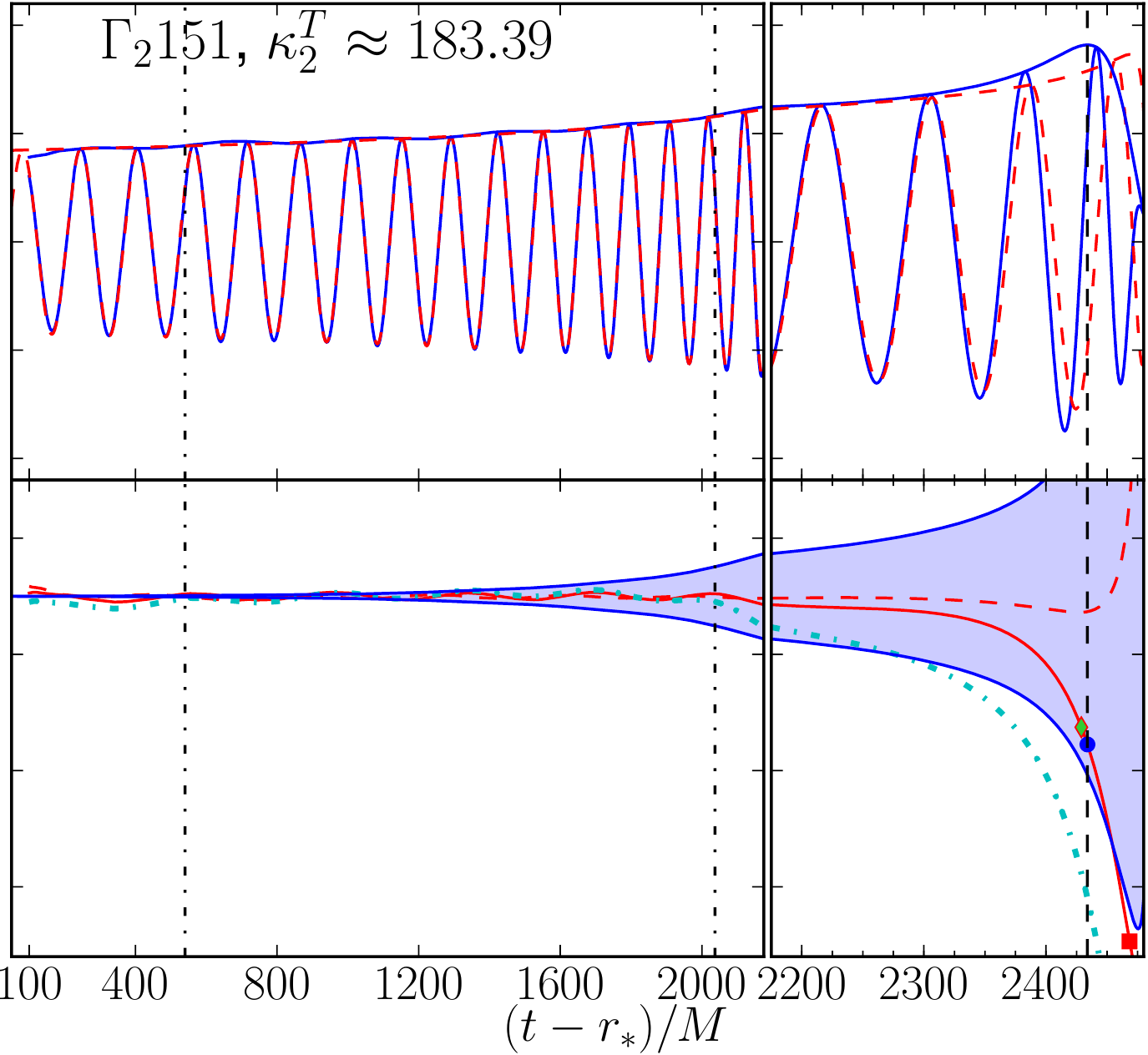}
     \caption{ 
      Phasing and amplitude comparison (versus NR retarded time)
      between \TEOBGSF, NR and the phasing of TT4 for three
      representative models.  
      Waves are aligned on a time window (vertical dot-dashed lines) 
      corresponding to  $I_\omega\approx (0.04,0.06)$. 
      The markers in the bottom panels indicate: the crossing of 
      the \TEOBGSF~LSO radius; NR (also with a dashed vertical line) 
      and EOB merger moments.}
    \label{fig:phasing}
  \end{center}
\end{figure*}

\paragraph{EOB-NR comparison: energetics.---}
We compare EOB to NR energetics using the gauge-invariant relation
between the binding energy and the orbital angular 
momentum~\cite{Damour:2011fu,Bernuzzi:2012ci,Bernuzzi:2013rza}. 
We work with corresponding dimensionless quantities defined 
respectively as $E_b \equiv \left[(M^0_{\rm ADM}
  - \Delta\mathcal{E}_{\rm rad})/M-1\right]/\nu$ and $j \equiv ({\cal
  J}^0_{\rm ADM} - \Delta\mathcal{J}_{\rm rad})/(M^2\nu)$, where
$\Delta\mathcal{E}_{\rm rad}$ ($\Delta \mathcal{J}_{\rm rad}$) is  
the radiated GW energy (angular momentum). Since the relation $E_b(j)$
essentially captures the conservative dynamics~\cite{Damour:2011fu},
this analysis directly probes the performance of the EOB Hamiltonian,
and notably the definition of $A^T(u;\nu)$. 

The top panels of  Fig.~\ref{fig:dynamics} compare for all EOS 
four energetics $E_b(j)$: NR, \TEOBGSF, \TEOBNNLO, and the PN-expanded tidal
energetics TPN, i.e. the (2PN accurate) expansion of the 
function $E_b(j)$ in powers of $1/c^2$.
The markers on the first three curves identify the corresponding merger
points. Following~\cite{Bernuzzi:2014kca}, we define the 
moments of merger, intrinsically for each model, 
as the peak of the modulus of the corresponding $\ell=m=2$ waveform.
The  two differences $\Delta E^{\rm EOBNR}_b(j)= E^{\rm EOB}_b(j)-E^{\rm NR}_b(j)$ 
for \TEOBGSF~and \TEOBNNLO~are shown in the bottom panels. 
The shaded area indicates the NR uncertainty. 
The main findings of this comparison 
are: 
(i)~TPN is always above the NR curve with a difference which
becomes unacceptably large towards merger (cf. the BBH
case in~\cite{Damour:2011fu});  
(ii)~the location of the 
\TEOBNNLO~merger point in the $(E_b,j)$ plane 
is, in all cases, very significantly away from the corresponding
NR merger point; 
(iii)~by contrast, the \TEOBGSF~merger point is,
in all but one case (2B), rather close to NR, especially when 
$\kappa_2^T$ is large;
(iv)~in all cases, the \TEOBGSF--NR 
differences (bottom panels) closely oscillate around zero during 
most of the simulated $\sim$ ten orbits;
(v)~moreover, such differences
keep staying within the NR uncertainty essentially up to 
(or slightly before for H4 and MS1b) the \TEOBGSF~merger.

\paragraph{EOB-NR comparison: phasing.---}
The EOB resummed tidal waveform is obtained 
following~\cite{Damour:2008gu,Damour:2012yf}.  
We compare the EOB and NR quadrupole waveforms ${\cal R} h_{22}$,
with ${\cal R}(h_+-{\rm i} h_\times)=\sum_{\ell m}{\cal R} h_{\ell
  m}\, {}_{-2}Y_{\ell m}$, by using a standard (time and phase) 
alignment procedure in the time domain.
Relative time and phase shifts are determined by minimizing
the $L^2$ distance between the EOB and NR phases integrated
on a time interval corresponding to the dimensionless frequency 
interval $I_\omega=M(\omega_L,\omega_R)= (0.04,0.06)$ for all EOS,
except $\Gamma_2164$, for which $I_\omega=(0.0428,0.06)$ as the simulation 
starts at higher GW frequency. 
Such choice for $I_\omega$ allows one to average out the phase 
oscillations linked to the residual 
eccentricity ($\sim 0.01$) of the NR simulations.

A sample of time-domain comparisons for three representative
$\kappa_2^T$'s is shown in Fig.~\ref{fig:phasing}.
Top panels compare 
the \TEOBGSF~and NR waveforms real part and modulus.
Bottom panels:
(i) phase and relative amplitude differences between \TEOBGSF~and NR; 
(ii) phase difference between the  tidal Taylor T4 
with NLO tides and 3PN waveform (TT4) and NR; and 
(iii) NR phase uncertainty (shaded region). The two vertical (dot-dashed)
lines indicate the alignment interval; as in Fig.~\ref{fig:dynamics},
the markers indicate the EOB (red) and NR (blue) mergers. 
The crossing of the radius of the \TEOBGSF~last 
stable orbit (LSO) is indicated by a green marker. 
The time-domain comparisons shows that for all $\kappa_2^T$ 
the \TEOBGSF~model is compatible with NR data {\it up to
  merger} within NR uncertainties (at the $2\sigma$ level or 
better, both in phase and amplitude). Note that the TT4 
phasing performs systematically worse than \TEOBGSF.

\begin{figure}[t]
  \begin{center}
    \includegraphics[width=0.45\textwidth]{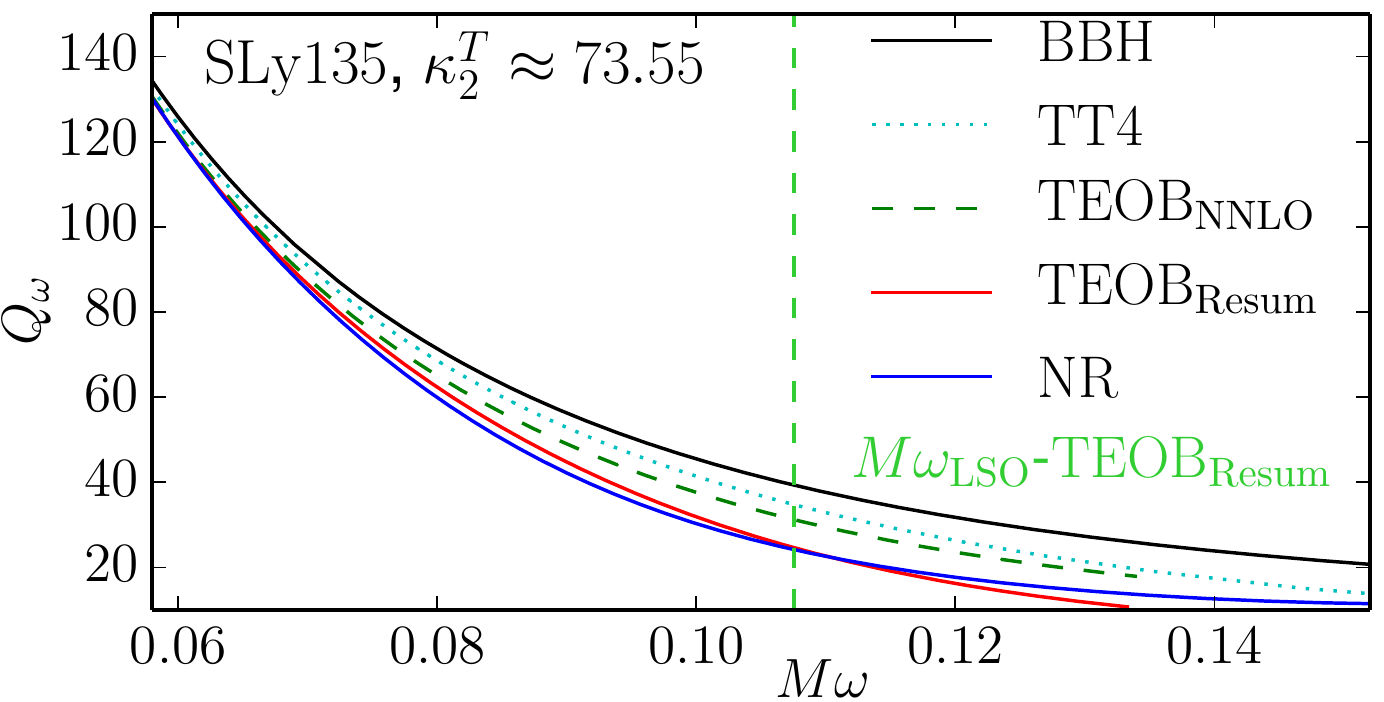}
    \caption{Phasing comparison of various analytical models 
      and with NR data using the gauge-invariant quantity 
      $Q_\omega\equiv \omega^2/\dot{\omega}$.}
    \label{fig:Qomg}
  \end{center}
\end{figure}

Figure~\ref{fig:phasing} is quantitatively completed
by Table~\ref{tab:configs}, which compares 
the phase differences 
$\Delta\phi^X\equiv \phi^{X}-\phi^{\rm NR}$ with $X={\rm
  TT4}$,\TEOBNNLO,\TEOBGSF~evaluated (after time-alignment) 
at the moment of NR merger. 
The NR uncertainty at merger $\delta\phi_{\rm NRmrg}^{\rm NR}$ is also listed 
in the table. These numbers indicate how the disagreement 
with NR systematically decreases when successively
considering the analytical models TT4, \TEOBNNLO~and \TEOBGSF.
Such hierarchy of qualities among analytical models 
is confirmed by the gauge-invariant phasing diagnostic
$Q_\omega(\omega)\equiv \omega^2/\dot{\omega}$~\cite{Baiotti:2010xh,Bernuzzi:2012ci}.
To clean up the eccentricity-driven oscillations in the NR phase,
we based our computation of $Q_\omega^{\rm NR}$ 
by starting from a simple, PN-inspired, six-parameter fit 
of the NR frequency as a rational function of $x=(\nu(t_c-t)/5+d^2)^{-1/8}$ 
(similarly to ~\cite{Hannam:2010ec}). For each $\kappa_2^T$ we find:
$Q_{\omega}^{\rm NR}\approx Q_{\omega}^{{\rm TEOB}_{\rm Resum}}<
Q_{\omega}^{{\rm TEOB}_{\rm NNLO}}< Q_{\omega}^{{\rm TT4}}<
Q_{\omega}^{\rm BBH}$ (see Fig.~\ref{fig:Qomg}, for SLy135).

\paragraph{Merger characteristics.---}
The \TEOBGSF~model, in addition to giving good
energetics, $E_b(j)$, and phasing $\phi(t)$ up to NR merger, has the 
remarkable feature of intrinsically {\it predicting} the frequency 
location and physical characteristics of merger in good quantitative 
agreement with NR results.  This can have important consequences for
building analytical GW templates. More precisely, the two 
quasiuniversal functional relations~\cite{Bernuzzi:2014kca} 
$E_b^{\rm mrg}(\kappa_2^T)$ and $M\omega^{\rm mrg}(\kappa_2^T)$ 
(as well as $j^{\rm mrg}(\kappa_2^T$) and the waveform amplitude 
at merger $A_{22}^{\rm mrg}(\kappa_2^T)\equiv |{\cal R} h_{22}^{\rm
  mrg}|(\kappa_2^T)$)  
predicted by \TEOBGSF~are close to the NR ones and significantly 
closer than those predicted by \TEOBNNLO~(while PN does not 
predict \textit{any} merger characteristic).
For $E_b^{\rm mrg}$ and $j^{\rm mrg}$ see Fig.~\ref{fig:dynamics}. 
For $M\omega^{\rm mrg}(\kappa_2^T)$,  the ratio 
$\omega_{\rm NR}^{\rm mrg}/\omega_{\rm TEOB_{\rm Resum}}^{\rm mrg}$
ranges from $1.06$ ($\Gamma_2164$) to $1.17$ (H4).
For $A_{22}^{\rm mrg}$, the ratio $A_{\rm 22\,NR}^{\rm
  mrg}/A_{22\,{\rm TEOB}_{\rm Resum}}^{\rm mrg}$ ranges  from 1.05
($\Gamma_2151$) to 1.15 (2B) (see also
Fig.~\ref{fig:phasing}). Finally, after  
alignment, the difference  $\Delta t_{\rm mrg}=t_{\rm mrg}^{\rm
  TEOB_{\rm Resum}}-t_{\rm mrg}^{\rm NR}$  
between EOB and NR merger times is only 
$\sim (-30M,\,-8M,\, -9M,\, +34M,\, +51M,\,+92M)$ for the six models. 
Such agreements are remarkable as no NR-tuning of the EOB 
waveform was performed.

\paragraph{Conclusions.---}
We introduced the first tidal EOB model able to describe the 
energetics and waveforms of coalescing BNS from the early inspiral 
{\it up to the moment of merger}. 
The EOB prediction for the binary dynamics as measured by the $E_b(j)$
curve agrees with NR data within their uncertainties for a sample of
EOS spanning a significant range of tidal parameters, Fig.~\ref{fig:dynamics}.  
The EOB and NR waveform phasing essentially agree within
the NR uncertainties up to the moment of merger.
This result is  a significant improvement with respect to 
previous work~\cite{Bernuzzi:2012ci,Hotokezaka:2013mm},
notably because no parameters were tuned.
Given the NR intrinsic uncertainties, and the possible
residual eccentricity influence, we refrain from further calibrating
the model at this stage. Once improved NR data 
will be available, we expect to be able to NR-inform the model,
e.g. by including next-to-quasi-circular corrections to the waveform.

 \begin{acknowledgments}
   \paragraph{Acknowledgments.---}  
  It is a pleasure to thank D.~Bini for sharing with us unpublished 
  results about the gravitomagnetic tides, and M.~Ujevic for providing 
  us with the NR initial data. 
  The EOB code developed here is publicly available at \verb1eob.ihes.fr1. 
  S.B. acknowledges partial support from the National
  Science Foundation under grant numbers NSF AST-1333520, PHY-1404569, and
  AST-1205732.  
  Ti.D. acknowledges partial support from the 
  DFG grant SFB/Transregio~7 ``Gravitational Wave Astronomy'' 
  and the Graduierten-Akademie Jena.
  Computations where performed on LRZ (Munich).
 \end{acknowledgments}

\bibliography{../../../Refs/sb_refs,inprep}{}

\end{document}